\RequirePackage{fix-cm}
\documentclass[natbib,smallextended]{svjour3}       
\smartqed  
\usepackage{graphicx}
\journalname{Acta Geodesica et Geophysica}
\begin{document}

\title{Greenhouse effect from the point of view of
radiative transfer
}


\author{S. Barcza}


\institute{   Geodetic and Geophysical Institute, Konkoly Observatory,
Research Centre for 
Astronomy and Earth Sciences, Hungarian 
Academy of Sciences, Sopron, Hungary \\
              \email{barcza@konkoly.hu}           
}

\date{Received: date / Accepted: date}

\maketitle

\begin{abstract}
Radiative power balance of a planet in the solar system 
is delineated. The terrestrial powers are transformed to average global flux
in an effective atmospheric column (EAC) approximation,
its components are delineated. The estimated and measured secular changes
of the average global flux are compared to the fluxes derived 
from the Stefan-Boltzmann law 
using the observed global annual temperatures in the decades between 1880 and 2010.
The conclusion
of this procedure is that the radiative contribution of the greenhouse gas
${\rm CO}_2$ 
is some 
$21\pm 7$~\% 
to the observed global warming from the end of the
XIXth century excluding the feedback mechanisms playing determining role
in the climate system.
Stationary radiative flux transfer is treated in an air column 
as a function of the column density of the absorbent. Upper and lower
limit of radiative forcing is given by assuming true absorption and
coherent scatter of the monochromatic radiation. An integral formula is given
for the outgoing long wave radiation (OLR) as a function of column density
of the components of greenhouse gases.
 \and Radiative fluxes 
\end{abstract}

\section{Introduction}
\label{intro}
Terrestrial atmospheric greenhouse effect, and its eventual change by the increment
of the concentration of greenhouse gases, resulting in
a global forcing, is a topic of many observational and theoretical papers. The 
present paper contributes to this field by presenting a solution to the 
problem of the greenhouse effect, adapted from the physics of stellar atmospheres
\citep{miha1}. The main difference in comparison to the stellar case
is that the conditions are far from local thermodynamic
equilibrium (LTE). The radiation has an asymmetry in 
the sense that the incoming downward radiation is in the optical domain
of wavelengths which is partly reflected, partly absorbed and converted to the emitted upward infrared 
radiation leaving finally the atmosphere. Furthermore, an appropriate time
averaging is necessary because the radiation field is variable in a planetary
atmosphere, there are diurnal and seasonal variations. 

The goal of the paper is to reveal long term 
changes in the radiative properties of a planetary atmosphere, 
which are significant for climatic considerations. The
elementary task is pursued here to solve the stationary radiative transfer 
problem in the case of an effective air column (EAC) or water column. The 
change of the
emerging flux is derived as a function of variable column density
of the greenhouse gases. This task is relatively simple,
it is a primary component in involved global climate models.
It can be separated from the 
problem to synthetize an overall atmospheric model which would treat
the hydrodynamical, radiative, chemical and phase changes and take into
account the geographical differences resulting finally in a climate
change. This coupled  
problem is beyond the scope of the present paper because
oceans, solid surface and atmosphere compose a very complex system 
and have a large number of degrees of freedom to store and emit the energy 
absorbed from the incoming flux. Furthermore, internal 
non-radiative degrees of freedom exist among which energy transfer takes 
place under conditions far from local thermodynamic equilibrium. These
factors can result in climate fluctuations by feedback mechanisms having
much larger characteristic times compared to the radiative processes
which are discussed in the present paper and result in changes over a
few days time scale.

Large amount of temperature measurements are available at the
terrestrial surface and in the atmosphere. Homogenizing these data and
to derive appropriate
global averages is an ambiguous task. Additional problem is
that the temperature is an intensive quantity of thermodynamics, that is,
temperatures are only statistical data 
for a quantitative measurement of the greenhouse effect. 
The flux and power are extensive quantities of thermodynamics and
atmospheric physics. The main goal of this paper to calculate 
temperatures from propagation, balance and conversion of the radiative flux. 
The discussions are restricted to radiative processes, the condition of
the convective instability is not discussed.
The crucial quantities are integrated power and flux of incoming and outgoing
electromagnetic radiation, the computation of specific intensity 
is not necessary. 

The power and flux describe the variable absorbed or emitted electromagnetic
radiation over time intervals extending from minutes to centuries.
The time interval must be sufficiently long 
if the greenhouse effect is to be treated. It must not be forgotten 
that the units of power and
flux, watts and watts/surface, rely for seconds, averaging these 
quantities for a whole planet and for time intervals century or longer must 
be done carefully.  

The paper is organized as follows. Section 2 defines the greenhouse
effect, introduces the powers and
fluxes which are the main quantities, the units
${\rm W}$
and
${\rm Wm}^{-2}$ 
are used for them. The
problem of converting them into the more familiar quantity of temperature
is indicated. Sec. 3 reviews the total solar irradiation and its variability.
Sec. 4 is devoted to derive the flux propagation, the connection of flux
with albedo and greenhouse effect is enlightened. The radiative contribution 
of the increasing concentration of the atmospheric greenhouse gases
for the present day global forcing is estimated in Sec. 5.  
The results are discussed in Sec. 6, the conclusions are drawn in Sec. 7. 

\section{The greenhouse effect, empirical approach}
\label{sec:2}
An empirical definition of the local greenhouse effect at a geographical
point was given by \citet{rava1} in terms of the flux
${\cal F}_{\rm G}$
\begin{equation}\label{2.101}
{\cal F}_{\rm G}={\cal F}_{\rm surf}-{\cal F}_{\rm OLR}
\end{equation}
where
${\cal F}_{\rm surf}$
and
${\cal F}_{\rm OLR}$
are the integrated flux emitted by the surface and the outgoing longwave flux
at the top of the atmosphere (TOA), respectively.
${\cal F}_{\rm G}$ 
is the trapped flux by the atmosphere, it warms the
atmosphere. Of course, these are 
integrated fluxes for the infrared wavelengths. 

The source of
${\cal F}_{\rm surf}$
is the total solar irradiation (TSI)
\begin{equation}\label{3.100}
{\cal F}_{\rm S}=1361.5\mbox{Wm}^{-2}d^{-2}
\end{equation}
\citep{tsi1} where
$d$
is the distance of the planet to the Sun in astronomical units
($1\:\mbox{au}=1.4962\times 10^8\:\mbox{km}$.) 
${\cal F}_{\rm S}$
enters in the atmosphere under different angles depending on geographical
position and rotation, 
it is disseminated over the whole Earth giving the averaged flux
${\cal F}_{\rm S}/4=340.38\:\mbox{Wm}^{-2}$.
This approximation treats the atmosphere as an effective air column (EAC). 
The theoretically computed fluxes of an EAC can be compared to 
averaged fluxes over different geographical locations if they 
were constructed by carefully averaging measured and
theoretically computed atmospheric
fluxes, for a part of the surface or for the whole Earth. The EAC model 
is applied for the whole Earth in many papers like e.g. \citep{tren2} etc. 
Definition ({\ref{2.101}) of
${\cal F}_{\rm G}$
must be supplemented by the contribution from the flux transfer 
between the optical irradiance and 
infrared flux which propagates in the atmosphere. 
The eventual infrared flux 
from the surface can easily be introduced in (\ref{2.101}) by an additive 
term to
${\cal F}_{\rm surf}$. 

The power 
$W_{\rm S}=\pi R^2{\cal F}_{\rm S}$
is available for a planet of radius
$R$
to cover the needed power of meteorological, climatological processes 
in the atmosphere, oceans and at the solid surface. 
$W_{\rm S}=1.723\times 10^{17}\:\mbox{W}$
is available for the whole Earth at
$d=1$.
Radiation is only possibility to exchange power with the interplanetary 
environment since the particle irradiation of the Sun is negligible in
comparison to that of the incoming electromagnetic radiation. The
power balance of the Earth (or a planet) is expressed at the top
of the atmosphere (TOA) by 
\begin{equation}\label{2.100}
W_{\rm S}=W_{\rm O}+W_{\rm OLR}-W_{\rm I}
\end{equation}
where 
$W_{\rm O}$,
$W_{\rm OLR}$
are the reflected, absorbed (and reemitted) powers in the optical and infrared 
wavelengths, respectively. 
$W_{\rm I}$
was added for the sake of completeness, it
is the power released by the Earth itself, e.g. from the radioactive internal 
heat production 
$4.42\times 10^{13}\;\mbox{W}$
\citep{lowr1}. It must be noted for orientation that the averaged power
release from the global industrial energy production 
$\approx 5.6\times 10^{20}\:\mbox{J/year}$
\citep{iea1} contributes to 
$W_{\rm I}$
$1.78\times 10^{13}\:\mbox{W}\approx 10^{-4}W_{\rm S}$.
However, this component of
$W_{\rm I}$ is negligible for the Earth in comparison to the
other powers in (\ref{2.100}), but the
$\mbox{O}(10^{-4}W_{\rm S})$
is a few percent of the net down value in the CERES period which
can be the suspected imbalance leading to the recent warming of the climate 
\citep{tren2}. 

The dissipation produces OLR from all meteorological, climatic processes. 
The greenhouse effect is stationary if all components are constant in 
(\ref{2.100}). Eq. (\ref{2.100}) takes the form in terms of time averaged 
integrated fluxes
\begin{equation}\label{2.102}
\textstyle{1\over 4}{\cal F}_{\rm S}
                   -{\cal F}_{\rm O}-{\cal F}_{\rm OLR}
                   +{\cal F}_{\rm I}=0
\end{equation}
at the TOA which is defined by the zero optical depth
$\tau=0$
for all wavelengths and 
${\cal F}_{\rm OLR}$
contains now the infrared flux converted from the optical one before reaching 
the surface.
Equation (\ref{2.102}) can be extended to any optical depth and it
renders possible to treat quantitatively the greenhouse effect in the 
EAC approximation. 

To determine the effect of the trapped flux
${\cal F}_{\rm G}$ 
on the climate is equivalent
to solve the problem of the propagation and absorption of radiation
in an air column and to convert the flux to local temperature by
taking into account the interaction of the radiation and atmospheric matter.
This interaction takes place among
conditions which are far from LTE. The
irradiation by the Sun at a planet differs considerably from the 
Planck-distribution of flux
\begin{equation}\label{2.105a}
{\cal F}_\lambda^{(+)}(T)={{2\pi hc^2}\over{\lambda^5(\exp\{hc/kT\lambda\}-1)}}
\end{equation} 
where
$T=T_{\rm e,\: Sun}=5780\:\mbox{K}$
and 
${\cal F}_\lambda^{(+)}(T)$
must be multiplied by the dilution factor
$d^{-2}$.
$^{(+)}$
indicates that the flux is emitted in the half hemisphere above the surface
of a celestial body. 
The emitted radiation of a terrestrial ocean follows 
the Planck-distribution within 1\% at infrared wavelengths 
(\citealt{rava1}, \citealt{beut1}) if  
$T\approx T_{\rm surf}$
is introduced in (\ref{2.105a}).
 
The inversion of the Stefan-Boltzmann law allows
to order an effective temperature to a flux
${\cal F}^{(+)}$:
\begin{equation}\label{2.105}
T_{\rm e}=\Bigl({{\cal F}^{(+)}\over{\sigma}}\Bigr)^{1/4}, \hbox{\ \ \ }
\sigma=5.672\times 10^{-8}{{\mbox{W}}\over{\mbox{m}^2\mbox{T}^4}}.
\end{equation} 
It must be emphasized that 
$T_{\rm e}$ 
is an effective, not a 
local temperature. It is rather a model parameter of the EAS because
the temperature in an air column is height dependent. The 
$T_{\rm e}$
at the top of the atmosphere for planets of the solar system and Saturn 
moon Titan is shown in Table \ref{tab:1} in comparison to
the empirically measured temperatures in the atmosphere or at the
surface of these celestial bodies. 
A strong forcing is obvious in the atmosphere of Venus which 
is much hotter than the temperature expected by the solar irradiation, 
in line with the Stefan-Boltzmann law. The atmosphere of the Earth and perhaps of
Titan produces a slight forcing while the forcing at the TOA of 
the other planets, -- that is the greenhouse effect -- is almost negligible.

\begin{table}
\caption{$T_{\rm e}=(\textstyle{1\over 4}{\cal F}_{\rm S}/\sigma)^{1/4}$ 
and empirical surface temperatures 
$T_{\rm surf}$ 
in the solar system}
\label{tab:1}       
\begin{tabular}{llll}
\hline\noalign{\smallskip}
 & $d$ [au] & $T_{\rm e}$ & $T_{\rm surf}$  \\
\noalign{\smallskip}\hline\noalign{\smallskip}
Venus & 0.72 & 328 & $\approx 730$  \\
Earth & 1 & 278 & $\approx 287$ \\
Mars & 1.52 & 226 & $\approx 218$  \\
Jupiter & 5.20 & 122 & $\approx 120^\ast$ \\
Saturn & 9.55 & 90 & $\approx 88^\ast$  \\
Titan & 9.55 & 90 & $\approx 95$  \\
\noalign{\smallskip}\hline
\end{tabular}
\begin{list}{}{}
\item[$^\ast$]: TOA temperatures
\end{list}
\end{table}

\section{The total solar irradiation (TSI)}
\label{sec:3}

The TSI($={\cal F}_{\rm S}$) 
is not constant. The observed variability originating from the Sun 
between 1979-2010 is reported
in \citep{tsi1}. The time scales of solar oscillations, solar magnetic cycle,
solar cycle and secular changes vary from 5 minutes to 
$10^5\mbox{-}10^6$
years. The amplitudes are known from observations over decade time
scale, they are below 
$10^{-3}{\cal F}_{\rm S}$  
for the oscillations and the cycles. The small downward drift of the observed TSI 
($<10^{-4}$
over decade time scale)
and the long term secular changes (which are known only from theoretical 
considerations with large uncertainty) seem to be irrelevant for the 
observed present day forcing (\citealt{tsi2}, upper panel in Fig.~\ref{fig:1}). 
For climatological
considerations the conclusion is that the averaged value 
of the TSI during the few decades of the observations can be regarded as 
constant. 

A critical remark is appropriate on the observed TSI. The ground based
observations resulted in
$1368d^{-2}\:\mbox{Wm}^{-2}$
a half century ago
\citep{unso1}. The results of Earth Radiation Budget Experiment (ERBE) 
and Clouds and the Earth`s Radiant Energy System (CERES) for the TSI are
$1367d^{-2}\:\mbox{Wm}^{-2}$
(1985-1989, minimal solar activity) and
$1365d^{-2}\:\mbox{Wm}^{-2}$,
(2000-2004, maximal solar activity), respectively.
The result of a Japanese reanalysis is 
$1356d^{-2}\:\mbox{Wm}^{-2}$,
while the average over 1979-2010 is
$1361.5d^{-2}\:\mbox{Wm}^{-2}$. 
An eventual explanation of the forcing by the TSI would require an ascending 
TSI 
$\approx 18\mbox{Wm}^{-2}$
(for 1880-2010) or
$\approx 12\mbox{Wm}^{-2}$
between 1979-2010, respectively. Changes of this size
are definitely not present in the observational results. 
This somewhat controversial situation is plotted in terms of
${\scriptstyle{1\over 4}}{\cal F}_{\rm S}$,
right part of Fig.~\ref{fig:1}, upper panel.

Two uncertain and hardly discussed components must be qualitatively mentioned  
here for the sake of completeness
which are connected with the celestial 
mechanics. The largest annual periodic variation
$\Delta{\cal F}_{\rm S}/{\cal F}_{\rm S}=\vert -2\Delta d/d\vert\approx 0.064$ 
($\Delta F_{\rm S}=87\:\mbox{Wm}^{-2}$)
is the consequence of the eccentricity 0.016 of the orbit of the Earth.
This exceeds the observed temporal variations of
${\cal F}_{\rm S}$
by almost two orders of magnitude.
Another less discussed component is the consequence of the tilt of the
rotation axis of the Earth with a precession period
$\approx 26000$ years. This effect is coupled to the precession of the orbit
with a period
$\approx 113000$
years. The periods are very close to a
$9:2$
resonance. These hardly discussed components can be coupled to possible climate
changes over $\approx$~230000 years
by the different absorbing capacity of the TSI in the Northern and 
Southern hemisphere because of the higher absorbing capacity of the oceans
\citep{gran1}.

\begin{figure*}
  \includegraphics[width=0.75\textwidth]{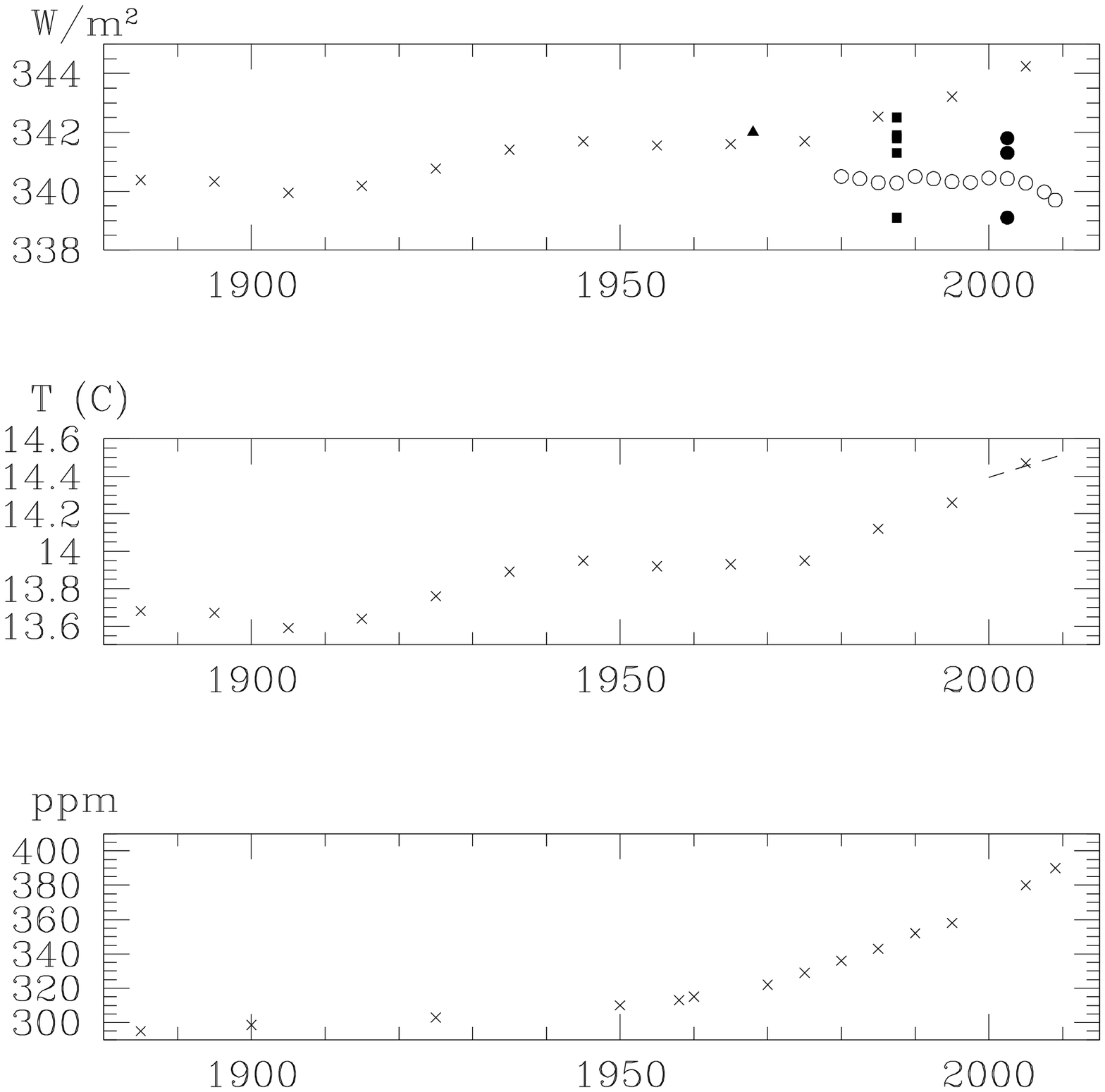}
\caption{{\bf Upper panel.} The fluxes
${\cal F}_{\rm S}/4$.
Crosses: 
$340.4+3\sigma\Delta T/T^3\mbox{Wm}^{-2},\: T=287\: \mbox{K}$
computed from the growth 
$\Delta T$
of the temperature in the middle panel.
Triangle: \citep{unso1},
filled squares and circles: fluxes from the observations of the
satellites ERBE and CERES, respectively, (the outlier points at the
bottom are from the Japanese reanalysis \citep{tren2},)
circles: fluxes from the satellites ACRYM etc. covering three solar cycles
(\citealt{tsi1}, \citealt{tsi2}). 
{\bf Middle panel.} Crosses: 
decadal global combined surface-air temperature (C)
over land and sea-surface, 1880-2010 \citep{jarr1},
these are the input data to the crosses in the upper panel.
Dashed line: linear regression to the annual averaged temperature, 2000-2010.
{\bf Lower panel}.
The atmospheric
$\mbox{CO}_2$
concentration (\citealt{jaco1}, \citealt{jarr1}.)
}
\label{fig:1}       
\end{figure*}

\section{Flux propagation in an air column, the radiative contribution
to the greenhouse effect}

The stationary transfer of specific intensity 
$I_\lambda(\theta ,\tau_\lambda)$
of wavelength
$\lambda$
is described in the EAC approximation by the
\begin{equation}\label{5.0.100}
{\rm cos} \theta{{{\rm d}I_\lambda(\theta ,\tau_\lambda)}
                 \over{{\rm d}\tau_\lambda}}=
I_\lambda(\theta ,\tau_\lambda)-S_\lambda(\theta ,\tau_\lambda)
\end{equation}
monochromatic transfer equation
where 
$\theta$
is the polar angle in a spherical reference system with axis
$z$
perpendicular to the surface,
$z=0$
at the bottom of the atmosphere (BOA). Isotropy is assumed in the
azimuth. The plan-parallel approximation
is involved in (\ref{5.0.100}), it
is permissible because of the small geometrical thickness of the atmosphere 
compared to the planetary radius. 
$S_\lambda(\theta ,\tau_\lambda)$ 
is the source function at the optical depth
\begin{equation}\label{5.0.101}
\tau_\lambda(z)=\tau_\lambda(z=0)-\int_0^z\mbox{d}z^\prime\kappa_\lambda(z^\prime),
\hbox{\ \ \ \ \ }{\rm d}\tau_\lambda=-\kappa_\lambda{\rm d}z,
\end{equation}
\citep{miha1},
(the units of
$I_\lambda$
and
$S_\lambda$
are 
$\mbox{Wm}^{-2}\mbox{sterad}^{-1}$),
$\kappa_\lambda=\kappa_{\lambda,{\rm at}}n(z)$
is the monochromatic absorption coefficient
which is the product of the photon absorbing or scattering cross section 
$\kappa_{\lambda,{\rm at}}$ 
and number density 
$n(z)$
of an atmospheric component at height
$z$,
their dimensions are
$\mbox{m}^2$
and
$\mbox{m}^{-3}$,
respectively. 
$\tau_\lambda=0$,
at the TOA. 

The line broadening in a planetary atmosphere produces a depth-dependent
Doppler-profile. It
can be approximated by a rectangle 
$P(\lambda)$
of Doppler-width belonging to a local or average temperature in the air column.
$P(\lambda)$
must be normalized to unity in the wavelength interval of the line,
$\kappa_{\lambda,{\rm at}}$
is an atomphysical quantity, it does not depend on geometrical height.
The optical depth at the TOA depends therefore only on the 
$N=\int\mbox{d}z^\prime n(z^\prime)$
column density of the absorbent:  
\begin{equation}\label{5.0.102}
\tau_\lambda=\kappa_{\lambda,{\rm at}}P(\lambda)N.
\end{equation}
(At the see level:
$N_{{\rm CO}_2}=5.47\times 10^{21}\;\mbox{cm}^{-2}$
for $400$ ppm concentration,
$N_{{\rm CH}_4}=6.80\times 10^{19}\;\mbox{cm}^{-2}$
for $1.808$ ppm.)

The simplification (\ref{5.0.102}) can be allowed because 
the spectral irradiation of the air
column is practically constant within the line profile. It is useful
because the target of this study is to compute the change  
of the emerging integrated flux at the TOA as a function of
$N$.
(The goals of the computation are neither the absolute value of the 
flux nor a line profile at the TOA.)

The intervals
$0\le\theta\le\pi/2$
and
$\pi/2\le\theta\le\pi$
are for the outward and inward fluxes
${\cal F}^{(+)\:{\rm ,}\:(-)}_\lambda(\tau_\lambda)
         =2\pi\int_{0\:{\rm ,}\:{\pi/2}}^{\pi/2\:{\rm ,}\:\pi}
          \mbox{d}\theta\sin\theta\cos\theta 
          I_\lambda(\theta,\tau_\lambda)$.
The Eddington approximation
can be applied for the upward and downward fluxes separately, that is
$\partial I^{(+)\:{\rm ,}\:(-)}_\lambda(\theta ,\tau_\lambda)/\partial\theta=0$
within the intervals 
$0\le\theta\le\pi/2$
and
$\pi/2\le\theta\le\pi$,
respectively, however,
${\cal F}^{(+)}_\lambda\not={\cal F}^{(-)}_\lambda$.
A multiplication of (\ref{5.0.100}) by
$2\pi\sin\theta\cos\theta$
and an integration over
$\theta\vert_0^{\pi/2}$
and
$\theta\vert_{\pi/2}^\pi$
leads to the equations 
\begin{eqnarray}\label{5.0.103}
{{\mbox{d}{\cal F}_\lambda^{(+)}}\over{\mbox{d}\tau_\lambda^\prime}}
   &=&{3\over 2}\Bigl[{\cal F}_\lambda^{(+)}-S_\lambda^{(+)}\Bigr],  \\
\label{5.0.104}
{{\mbox{d}{\cal F}_\lambda^{(-)}}\over{\mbox{d}\tau_\lambda^\prime}}
   &=&-{3\over 2}\Bigl[{\cal F}_\lambda^{(-)}-S_\lambda^{(-)}\Bigr].
\end{eqnarray}
describing the propagation of flux in the upward and downward directions. 

The first term on the right hand side of (\ref{5.0.103}) and (\ref{5.0.104})
accounts for the monochromatic absorption, the source functions
$S_\lambda^{(+)}$,
$S_\lambda^{(-)}$
describe how the absorbed radiation is reemitted. There are three relevant
elementary processes for the reemission. 

\paragraph{Coherent scatter}
is an elastic scatter of the photons from the microphysical point of view,
the wavelength of the absorbed and emitted photon is identical.
The source function is in this case
$S_\lambda^{(+)\:\mbox{,}\:(-)}={\cal F}_\lambda^{(+)\:\mbox{,}\:(-)}/2$. The 
upward monochromatic flux is at optical depth
$\tau_\lambda\le\tau_\lambda^\prime\le 0$ 
\begin{equation}\label{5.0.105}
{\cal F}_\lambda^{(+)}(\tau^\prime_\lambda)
     ={\cal F}_\lambda^{(+)}(\tau_\lambda)
     \exp\{-3(\tau_\lambda-\tau_\lambda^\prime)/4\}
\end{equation}
from the integration of (\ref{5.0.103}) where
${\cal F}_\lambda^{(+)}(\tau_\lambda)$
is the (infrared) upward flux at the BOA, that is at
$\tau_\lambda(z=0)$.
A downward (infrared) flux 
$1-{\cal F}_\lambda^{(+)}(\tau^\prime_\lambda)$
is present in the EAC. Their sum is constant. These formulae describe
the inward (optical) flux if 
$^{(+)}$
is replaced by
$^{(-)}$
and
${\cal F}_\lambda^{(-)}(\tau_\lambda=0)$
denotes the flux of the spectral solar irradiance (SSI) at the TOA. The incoherent
scatter \citep{miha1} can formally be merged in the coherent scatter, it
needs no separate treatment from the point of view of flux propagation.

\paragraph{Extinction of the photons}
of wavelength
$\lambda$
is expressed in terms of source function by
$S_\lambda^{(+)\:\mbox{,}\:(-)}=0$:
the absorbed part from the monochromatic flux
${\cal F}_\lambda$  
vanishes, it is transferred into another degree of freedom, e.g. to 
macroscopic or microscopic kinetic energy (winds or thermalization etc.)
or the absorbed flux of wavelength
$\lambda$
appears at another wavelength 
$\lambda^\prime$.
This latter process can be expressed mathematically by 
$S_{\lambda^\prime}^{(+)\:\mbox{,}\:(-)}\not=0$.

The extinction is important in the optical wavelengths
when optical flux is converted to infrared one. The flux from the
SSI is at
$\tau_\lambda^\prime$
\begin{equation}\label{5.0.106}
{\cal F}_\lambda^{(-)}(\tau^\prime_\lambda)
     ={\cal F}_\lambda^{(-)}(\tau_\lambda=0)
     \exp\{-3\tau_\lambda^\prime/2\}.
\end{equation}

\paragraph{True absorption} is a process in a planetary atmosphere if the gas and 
the radiation are in approximate equilibrium, that is, the conditions are close
to LTE. This holds for the upward flux in the infrared domain: the 
monochromatic infrared irradiation 
${\cal F}^{(+)}(\tau_\lambda)$
is approximately the Planck-distribution (\ref{2.105a}) at BOA 
\citep{beut1} and (\ref{2.105a}) can be assumed above the BOA for the whole 
interval
$z(\tau_\lambda^\prime),\:0\le\tau_\lambda^\prime\le\tau_\lambda$
with local temperature
$T(z)$.
This is expressed by using (\ref{2.105a}) in
$S_\lambda(z)=-S_\lambda(z)={\cal F}^{(+)}_\lambda[T(z)]$.
The local temperature is equal in first approximation
to the local effective temperature of the irradiating flux:
$T(z)=T_{\rm e}(z)$,
$+\mbox{corrections}$.

\subsection{The OLR}
The OLR is the integral over the appropriate infrared wavelengths:
\begin{equation}\label{5.1.100}
{\cal F}^{(+)}(\tau_\lambda^\prime=0)
     =\int\mbox{d}\lambda{\cal F}_\lambda^{(+)}(\tau_\lambda)
     \exp\{-K_\lambda\}
\end{equation}
where 
$K_\lambda=k\tau_\lambda=k\kappa_{\lambda,{\rm at}}P(\lambda)N$
and
$k={\textstyle{3\over 4}}$
or
$k={\textstyle{3\over 2}}$
are for the coherent scatter or extinction, respectively.
The atmosphere is optically thin in a line if
$K_\lambda\ll 1$. 
Remarkable is that 
$N$
is common for an atmospheric component at any
$\lambda$
and the OLR linearly depends on
$N$
if
$K_\lambda\ll 1$. 

The empirical definition (\ref{2.101}) of the greenhouse effect is expressed
by
\begin{equation}\label{5.1.101}
{\cal F}_{\rm G}=\int\mbox{d}\lambda[{\cal F}^{(+)}(\tau_\lambda)
                 -{\cal F}^{(+)}(\tau_\lambda^\prime=0)]
                =\int\mbox{d}\lambda{\cal F}^{(+)}(\tau_\lambda)
                 [1-\exp\{-K_\lambda\}],
\end{equation}
the integration is over the infrared wavelengths. The depth dependent emittance
(conversion of optical irradiation to infrared emittance, an additive term) 
was neglected in (\ref{5.1.101}).

\subsection{The albedo} 
The albedo is obtained by an integration of the reflected
flux (\ref{5.0.105}) from the coherent scatter
\begin{equation}\label{5.2.100}
a={{\int\mbox{d}\lambda{\cal F}_\lambda^{(-)}(\tau_\lambda=0)
     \Bigl[1-\exp\{-3\tau_\lambda/4\}\Bigr]}
     \over{\int\mbox{d}\lambda{\cal F}_\lambda^{(-)}(\tau_\lambda=0)}}.
\end{equation} 
(The integration is over the optical wavelengths.)

\subsection{The absorbed radiation}

The absorbed upward flux is
\begin{equation}\label{5.3.100}
{\cal F}^{(+)}(z^\prime)=\int\mbox{d}\lambda
               {\cal F}_\lambda^{(+)}(z=0)
               \Bigl[1-\exp\Bigl\{K_\lambda-k\kappa_{\lambda,{\rm at}}P(\lambda)
                \int_0^{z^\prime}\mbox{d}zn(z)\Bigr\}\Bigr]
\end{equation}
is available to forcing of the atmosphere at a height
$z^\prime$. 
It is maximal if extinction dominates 
the interaction of the radiation and atmospheric gases
($k={\scriptstyle{3\over 2}}$), 
this is expected in optically thick atmospheres, e.g. in Venus. It is minimal
if coherent scatter dominates
($k={\scriptstyle{1\over 2}}$), 
that is in optically thin atmospheres. The rule is for an EAC:
minimal true absorption-maximal OLR and
maximal true absorption-minimal OLR.
The terrestrial atmosphere is between these two limits, it is
approximately optically thin in optical wavelengths while it is 
thick in some infrared wavelengths where true absorption is expected to 
dominate the radiation transfer (e.g. at
$\approx 670\pm 100\:\mbox{cm}^{-1}$
or
$\approx 1300\:\mbox{cm}^{-1}$
for
$\mbox{CO}_2$
and
$\mbox{CH}_4$,
respectively.)
 
\subsection{Remark}

For actual computations, it is useful to remark that Taylor expansions
can easily be constructed for the monochromatic functions 
(\ref{5.0.105})-(\ref{5.2.100}) in terms of
$\tau_\lambda$.
These expansions can be transformed to expansions in terms of the
$\Delta N$
variation of the column density of an absorbing gas. 
$\Delta N$ 
is common for all
$\lambda$.

\section{The radiative role of
$\mbox{CO}_2$ 
in the change of the greenhouse effect}
\label{sec:4}

The clear sky forcing was measured by the Atmospheric
Emitted Radiance Interferometer (AERI) in the decade 2000-2010  \citep{feld1}: 
an average growth 
$\Delta{\cal F}=0.2\pm 0.065\:\mbox{Wm}^{-2}$
has been found with large seasonal ranges
$0.1\mbox{-}0.2\:\mbox{Wm}^{-2}$.
It was assumed that this growth under the cloudless sky can 
be attributed exclusively to the increase 
\begin{equation}\label{5.0.200}
\Delta\nu=\Delta N/N=22\:\mbox{ppm}
\end{equation}
of the 
$\mbox{CO}_2$ 
concentration during this period.
This observational result can be used in the EAC approximation to
estimate the contribution of the increasing
$\mbox{CO}_2$
concentration to the global present day forcing.
The increment of the flux from the AERI observations must be compared to that 
from the Stefan-Boltzmann law if the observed increase 
$\Delta T$
of the temperature at the surface (middle panel in Fig. \ref{fig:1})
is substituted for
$\Delta T_{\rm e}$
in
\begin{equation}\label{4.1.100}
\Delta{\cal F}={{4\Delta T_{\rm e}}\over{T_{\rm e}}}{\cal F}.
\end{equation}

\begin{itemize}
\item[$\bullet$]
The increment is
$\Delta T=0.124\pm 0.061$
from a linear regression for the decade 2000-2010 (dashed line in the middle
panel of Fig. \ref{fig:1}.) The growth of flux is
$\Delta{\cal F}=0.66\pm 0.32\:\mbox{Wm}^{-2}$
from (\ref{4.1.100}) if the BOA values
$T_{\rm e}=287\:\mbox{K}$,
${\cal F}=384\:\mbox{Wm}^{-2}$ 
are used. Dividing the increments gives that
$30\pm 15$ 
percent of the growth originates from the increasing atmospheric 
concentration of
$\mbox{CO}_2$.
($\Delta{\cal F}=0.58\:\mbox{Wm}^{-2}$
if the more recent
${\cal F}=320\pm 10\:\mbox{Wcm}^{-2}$
\citep{step1} is used in (\ref{4.1.100}) resulting in the slightly higher 
35\% contribution of 
$\mbox{CO}_2$.)

\item[$\bullet$]
An improved estimation can be obtained if the flux growth observed by 
AERI is extrapolated for the longer period 
$\approx 1750\:\mbox{to}\:2010$.
The increment is 
$290\rightarrow 400\:\mbox{ppm}$
($\Delta\nu=+110\:\mbox{ppm}$) 
in the atmospheric concentration of
$\mbox{CO}_2$
from the beginning of the industrial revolution in the
XIXth century: the extrapolated result is
$\Delta{\cal F}=1.0\pm 0.33\:\mbox{Wm}^{-2}$.
The secular increment of temperature at the BOA 
(Fig. \ref{fig:1}, crosses in the middle panel) is
$\Delta T_{\rm e}=0.88\:\mbox{C}$
yielding
$\Delta{\cal F}=4.72\:\mbox{Wm}^{-2}$
from (\ref{4.1.100}).
$(1\pm 0.32)/4.72=21\pm 7\;\mbox{percent}$
is the contribution of the
$\mbox{CO}_2$ to the radiative forcing from the 
$\Delta\nu\approx 110\:\mbox{ppm}$
increment. 
\end{itemize}

\section{Discussion}

\subsection{Conversion of flux to temperature}
Temperature is an important quantity for the meteorology or climatology.
It can be measured easily, large data sets are available at a great number 
of geographical points on the land. The homogenization and derivation
of global temperature averages 
form a non-trivial task. Relevant quantity is the flux
for a physical interpretation of climatic processes. Conversion between temperature
and flux can be done if the interaction of flux (power/surface)
with planetary surface and atmosphere is described properly: the 
average energy
density of the planetary matter is calculated from the 
balance of the input and output power density.
The factors depend on geographical position and diurnal variation.

The effective temperature 
$T_{\rm e}$
of the radiation can be derived from the flux by the Stefan-Boltzmann law,
however, the interaction of the radiation and matter takes place 
under conditions that are far from LTE.
Only a part of the radiative flux is converted to internal 
energy of the gas to which temperature can finally be attached unequivocally. Eq. 
(\ref{2.100}) is a global power balance, dividing it by the terrestrial surface
enables to convert it to flux data and to use the EAC approximation. 

Comparing the observed flux and temperature data for 1980-2010  
(Fig.~\ref{fig:1}) reveal
that the variation of the TSI is excluded as a cause for the increase of the
global temperature in the XXth century.
 
The concentration of
$\mbox{CO}_2$
is monotonically increasing in the XXth century (Fig.~\ref{fig:1}, lower 
panel) while 
$\Delta T$ 
shows undulatory behaviour with a characteristic time $\approx$~40 years: the 
rise between 1905-1935 is followed by the stagnation 
$\Delta T\approx 0$ 
between 1935-1975 and a more intensive rise from 1975. The slope is 
$\approx 0.157\:\mbox{C/decade}$
between 1980-2000, while it decreased to
$0.124\pm 0.061\:\mbox{C/decade}$
in 2001-2010 (leading to speculations where is the missing heat, 
\citealt{toll1}.) A resolution of this problem can be that the storing capacity
of the atmosphere follows the radiative input with some delay, e. g.
because of the role of the heat reservoirs represented by the solid
surface and oceans. These have a much larger capacity for storing heat than
the atmosphere merely. Another factor for the non-simultaneous behaviour is that
the radiation and the atmosphere are not in local thermodynamic equilibrium (LTE,
\citealt{miha1}.)  
  
A hardly examined radiative factor of the non-concordance of TSI and global 
warming can 
be the spectral difference of the atmospheric and black-body radiation: the
derived global combined surface-air temperature is not the effective
temperature:
$T(z=0)\not=T_{\rm e}(z=0)$.
However, 
$\Delta T_{\rm e}=\Delta T$ 
is a more reliable input in the 
Stefan-Boltzmann law as an estimation of first step because its basis, the 
Planck spectral distribution of the radiation is a good 
first approximation for emission into the atmosphere. An argument directing 
attention to this 
problem is that the proper zero point of the crosses in the upper
panel would be
$384.8\:\mbox{Wm}^{-2}$
belonging to
$T=287\:\mbox{K}$
at the BOA \citep{jarr1}, not the applied
$340.4\:\mbox{Wm}^{-2}$
belonging to the lower
$T_{\rm e}=278.3\:\mbox{K}$
derived from the TSI at the TOA (Table \ref{tab:1}) or the even lower
$\approx 320\:\mbox{Wm}^{-2}$ \citep{step1}. 

Nevertheless, a much better approximation can be expected for the secular change
$\Delta{\cal F}$
if merely the changes
$\Delta T$
from the global combined surface-air temperature curves 
(Fig. \ref{fig:1}) are used in the EAC approximation as
$\Delta T_{\rm e}$
in (\ref{4.1.100}).
A comparison of  
$\Delta{\cal F}$
derived from 
$\Delta T(z=0)=0.88\pm 0.1\;\mbox{C}$
and the AERI measurements is permissible since the changes could be estimated with
better relevance even if the zero point of
${\cal F}$
was less certain. The result of this comparison is that
$30\pm 15$ percent
has been found for the contribution of increasing
$\mbox{CO}_2$
concentration under clear-sky conditions. 
(It must be remarked that the reported clear-sky measurements with the AERI 
merge the effect of
$\mbox{CO}_2$
and the other greenhouse gases (e.g. 
$\mbox{CH}_4$,
$\mbox{N}_2\mbox{O}$),
etc. \citep{jaco1},
their concentration changed during the period 2000-2010 as well as that
of
$\mbox{CO}_2$
\citealt{jarr1}). 

The more reliable percentage
$21\pm 7$
as the effect of human industrial activities since $\approx$1750 is obtained
if the the observed
$\Delta{\cal F}$
of the AERI measurements is extrapolated for this whole period 
and it is compared to the temperature data for 120 years \citep{jarr1}.
(The error 7\% was derived from the error of the flux measurement. The
inclusion of the estimated error of
$T_{\rm e}$
would increase it to $\approx 9-10$~\%.)
This extrapolation is appropriate if the atmosphere is optically thin for
the greenhouse gases, 
that is the higher terms of the expansion (\ref{5.1.100}) are negligible:
$K_\lambda\ll 1$
for all relevant infrared wavelengths. This condition would only 
be satisfied if
$\kappa_{\lambda,{\rm at}}\ll 5\times 10^{-19}\:\mbox{cm}^2$ 
for the photon scattering cross section of the infrared transitions
in the spectrum 
($\lambda=5\mbox{-}50\:\mbox{micron}$)
of
$\mbox{CO}_2$
as follows from (\ref{5.0.102}). The condition is not satisfied
at see level
e.g. in the centre of the $\mbox{CO}_2$ bands ($\approx 16\:\mbox{micron}$.) 
However, the obtained 21\% is a good starting point to more refined
computations because the Taylor expansion of the fluxes in terms of
$\Delta\nu$
($=\Delta N/N$)
converges better.

\citet{rava1} estimated that the greenhouse effect of a doubling of 
$\mbox{CO}_2$
is
$\Delta{\cal F}=4\:\mbox{Wm}^{-2}$
($+16\:\mbox{W}$
in TSI) giving the averaged slope 
$\Delta {\cal F}/\Delta\nu =1.38\times 10^{-2}\:\mbox{Wm}^{-2}/\mbox{ppm}$
for the whole period from 1750. 
More recent radiative transfer models involving better data of the radiative 
properties of
$\mbox{CO}_2$
gave the slope of the global mean radiative forcing 
$\Delta{\cal F}/\Delta\nu=(1.66\pm 0.17)\times 10^{-2}\:\mbox{Wm}^{-2}/\mbox{ppm}$
at the tropopause \citep{feld1} as a consequence of the growth of
$\mbox{CO}_2$
concentration since 1750. Comparing these data to the increase 
of flux derived from the secular increase of temperature at the BOA 
(Fig. \ref{fig:1}, middle panel) gives 
that some
$\approx 21\pm 7$
percent is the contribution of the
$\mbox{CO}_2$
to the radiative forcing for the 
$\Delta\nu\approx 110\:\mbox{ppm}$
rise. However, this is significantly higher than
$\Delta{\cal F}/\Delta\nu=(9.09\pm 0.27)\times 10^{-3}\:\mbox{Wm}^{-2}/\mbox{ppm}$
extrapolated from the AERI measurements. 
The discrepancy suggests that the 
theoretically computed increment of the flux from the rising concentration
of the greenhouse gase
$\mbox{CO}_2$
may need a revision.

The formulae (\ref{5.1.100}) and (\ref{5.2.100}) 
offer an opportunity to estimate the quantitative dependence
of greenhouse forcing as a function of the (change) of the atmospheric
components. Remarkable is the primary dependence on column density and
the almost perfect independence on the temperature represented by
$P(\lambda)$
in the EAC. 

\section{Conclusions}

In EAC approximation,
the empirical definition of the atmospheric greenhouse effect has been 
formulated in fluxes. Flux is extensive
quantity in the theory of the radiative transfer. Problems to convert
the fluxes to more accessible intensive quantity temperature have only
been sketched.

A comparison of observed forcing between 2000-2010 and
the integrated fluxes from the terrestrial surface temperatures
derived by the Stefan-Boltz\-mann-law has shown that the rise of the
atmospheric concentration of
$\mbox{CO}_2$
from the start of the industrial revolution has
contributed to the radiative forcing of the atmosphere by some
$21\pm 7$
percent. This percentage cannot exceed 39 percent even if the most
recent LBL calculations are applied to compute the forcing, however, they
probably need a revision to be consistent with the AERI observations. The 
discrepancy between the 
percentage of forcing from the empirical basis and the theoretical
calculations indicates a need to reconsider the theoretical calculations
concerning the radiative role of the increasing
$\mbox{CO}_2$
concentration. A formalism of this has been described in this paper. 

The main result has been that the radiative forcing from the growth of
concentration of the non-condensing greenhouse gas
${\rm CO}_2$, 
released by the industrial 
activity in the XXth century, contributed to the observed forcing up
to $\approx 20-30$~\%. The presented results from treating the radiative
flux propagation in the atmosphere have left open the question for
the source of the remnant global forcing $\approx 70-80$~\%. On
speculation level the conclusion is probable that
the relaxation times of the interaction of radiation
and atmospheric gases under non-LTE conditions, the
effect of the aerosols (produced by industrial
activity and natural processes) and water (in non-condensed or condensed form)
seem to be more important regulators of the observed global warming in 
the XXth century. Furthermore, long term power release of accumulated heat from 
the oceans and solid surface can contribute to the $\approx 70-80$~\%.


%
\begin{acknowledgements}
The author is grateful to D. Drahos, L. Haszpra, L. van Driel-Gesztelyi 
I. K\'esm\'arky for many discussions and technical help. Comments of
anonymous referees are gratefully acknowledged.
\end{acknowledgements}
\bibliographystyle{aps-nameyear}      
\bibliography{example}                

\begin{thebibliography}{}
%

\bibitem[Beuttner \& Kern(1965)]{beut1}
Beuttner K. J. K., Kern C. D., 1965, J. geophys. Res. 70, 1329

\bibitem[Feldman et al.(2015)]{feld1}
Feldman D. R., Collins W. D., Gero P. J., Torn M. S., Mlawer E. J.
\& Shippert T. R., 2015, Nature 519, 339

\bibitem[Granopolski et al.(2016))]{gran1}
Granopolski A., Winkelmann R., Schellnhuber H. J., 2016, Nature 529, 200 

\bibitem[IEA(2014)]{iea1}
International Energy Agency Report, 2014 

\bibitem[Jacob(1999)]{jaco1}
Introduction to Atmospheric Chemistry, 1999, Princeton Univ. Press 

\bibitem[Jarraud(2013)]{jarr1}
Jarraud M., 2013
World Meteorological Organization, WMO-No. 1119

\bibitem[Lowrie(2007)]{lowr1}
Lowrie W. 2007, Fundamentals of Geophysics, Second Ed., Cambridge Univ. Press

\bibitem[Mihalas(1978)]{miha1}
Mihalas D. 1978, Stellar Atmospheres, San Francisco, W. H. Freeman and Co.

\bibitem[Raval \& Ramanathan(1989)]{rava1}
Raval A., Ramanathan V., 1989, Nature 342, 758 

\bibitem[Solanki \& Unruh(2013)]{tsi1}
Solanki S. K., Unruh Y. C., 2013, Astr. Nachr. 334, 145

\bibitem[Solanki et al.(2013)]{tsi2}
Solanki Sami K, Krivova Natalie A, Haigh Johanna D., 2013, ARA\&A 51, 311

\bibitem[Stephens et al.(2012)]{step1}
Stephens G. L. et al. 2012, Nature Geoscience 5, 691

\bibitem[Tollefson(2014)]{toll1}
Tollefson J., 2014, Nature 505, 276

\bibitem[Trenberth, Fasullo \& Balmaseda(2014)]{tren1}
Trenberth K. E., Fasullo J. T., Balmaseda M. A., 2014, J. of 
Climate. 27, 3129 DOI:10.1175/JCLI-D-00294.1

\bibitem[Trenberth, Fasullo \& Kiehl(2009)]{tren2}
Trenberth K. E., Fasullo J. T., Kiehl J., 2009, Bull. Amer. 
Meteor. Soc. 90, 311, DOI:11.1175/2008BAMS2634.1

\bibitem[Uns\"old(1968)]{unso1}
Uns\"old A., 1968, Physik der Sternatmosph\"aren, Springer

\end{thebibliography}
\nocite{*}


\end{document}